# On the Evolution of Phenomenal Consciousness

*Evolution de la conscience phénoménale*

Jean-Louis Dessalles*    and    Tiziana Zalla** (1)

*   ENST -   Département Informatique & Réseaux
*46 rue Barrault - 75013 Paris - France*
*dessalles@enst.fr*

**   CREA -   Ecole Polytechnique
*1 rue Descartes - 75005 Paris - France*
*zalla@poly.polytechnique.fr*

A number of concepts are included in the term "consciousness". We choose to concentrate here on phenomenal consciousness, the process through which we are able to experience aspects of our environment or of our physical state. We probably share this aspect of consciousness with many animals which, like us, feel pain or pleasure and experience colours, sounds, flavours, etc. Since phenomenal consciousness is a feature of some living species, we should be able to account for it in terms of natural selection. Does it have an adaptive function, or is it an epiphenomenon ? We shall give arguments to reject the second alternative. We propose that phenomenal properties of consciousness are involved in a labelling process that allows us to discriminate and to evaluate mental representations. We also discuss to what extent consciousness as such has been selected for this labelling function.

*Le terme de "conscience" recouvre plusieurs concepts. Nous parlons ici de conscience phénoménale, cet ensemble de processus par lesquels nous avons une expérience de certains aspects de notre environnement et de notre état physiologique. Nous partageons probablement cet aspect de la conscience avec de nombreuses espèces animales qui, comme nous, ressentent de la douleur et du plaisir, et font l'expérience des couleurs, des sons, des odeurs, etc. Comme la conscience phénoménale est une caractéristique de beaucoup d'espèces vivantes, nous devons l'expliquer en invoquant la sélection naturelle. A-t-elle une fonction adaptative, ou est-elle un simple épiphénomène ? Nous donnons des arguments en faveur de la première option. Nous suggérons le fait que les propriétés phénoménales de la conscience sont impliquées dans un processus d'étiquetage qui nous permet de discriminer et d'évaluer les représentations mentales. Nous discutons ensuite l'hypothèse selon laquelle la conscience en tant que telle aurait été sélectionnée pour cette fonction d'étiquetage.*

---

[1] Authors' names are listed alphabetically.                                                                  oct. 1996

# 1. Introduction

The term "consciousness" is merely a label for many separate phenomena. Whatever consciousness is, it is something that, far from being an undifferentiated stream of inner events, is instead a composite phenomenon which corresponds to the activity of functionally differentiated modular systems. We need to distinguish here *phenomenal consciousness* from other cognitive processes, from conceptual knowledge and from higher-order conscious states [Block 1995]. Phenomenal consciousness refers to qualitative properties of experience. The vividness of pain, pleasure, redness, the taste of red wine are examples of qualitative experiences[1]. We want to deal here with phenomenal consciousness for two main reasons. First, many non-human species may have qualitative experience (if we think of pain). We do not need to grant them other aspects of consciousness [Griffin 1981], just the ability to experience smell, colour or pleasure, even if their experience is qualitatively different from ours. An obvious question then arises : when, and why, did qualitative experience arise in phylogenetic history ? Our second motivation for considering phenomenal consciousness comes from the fact that its very existence, and the biological adaptive function it fulfils, remain highly mysterious. Is it a mere epiphenomenon, or worse : a non-scientific object, or does it play a precise, essential, biological role ?

In our view, in order for the concept of consciousness to be of scientific interest, one has to show that it is a natural kind, *i.e.* a phenomenon which is useful and convenient to isolate as *explanandum* for a more advanced scientific theory. In this presentation, we take the existence of phenomenal consciousness as granted and we attempt to make this existence compatible with evolutionary principles. We shall study first what properties characterise qualitative experience, before looking for a possible adaptive function.

We shall first consider the possibility that phenomenal consciousness is a mere fortuitous epiphenomenon, and that complex cognitive behaviour can take place in its absence. However we will reject this possibility. We will consider a possible account for phenomenal consciousness : it will be presented as a way to *label* experiences and mental states. Labelling is an essential feature of cognitive processing, though the most obvious way of labelling information, as used in computers, is not plausible in a neural implementation. Labelling through synchronous binding, and its qualitative correlate, will be presented as a solution which evolved to cope with environmental and behavioural complexity. Lastly, we will observe that this account, expressed first in terms of physical neural states, is insufficient to predict characteristic aspects of phenomenal consciousness. We will suggest that the ability to have experience is part of our phenotype and was retained as such by natural selection.

# 2. Properties and role of phenomenal consciousness

*2.1 Is conscious experience nothing but an evolutionary epiphenomenon ?*

What use is the ability to experience mental states or events in the outer world ? Most of the complex processes going on in our body are achieved without involving any conscious component. We are not conscious of our immune system, we do not feel each contraction of our stomach, we are not aware of maintaining our equilibrium at each moment. Many of our cognitive processes are performed without us being aware of them. We are even unable to monitor such processes. The way we analyse a visual scene, the way we recognise words in a complex acoustic signal, the way we adapt our walk on an uneven ground are good examples of such processes. In Fodor's terms [Fodor 1983], these processes are achieved by domain specific *modules* which are characterised by their encapsulation and their relative inaccessibility. Only the output, in the case of sensory modules, is experienced by the living being. Such unconscious processes may be quite complex, they may be context sensitive (for instance priming may affect word recognition) even if, according to Fodor, modules are encapsulated and thus receive little influence from other processes. If we think of complex unconscious tasks like shape and object recognition, we may wonder why cognition involves consciousness at all. Why are we sentient beings, why are we not unconscious like robots ? From a Darwinian perspective, this is a crucial question : what is the adaptive value of consciousness, and if consciousness has no such value, why do we happen to be conscious ?

This question is even more vital from the functionalist perspective. Functionalism considers that what is relevant in cognition is the causal network of mental states which is involved in cognitive computations. Consciousness plays no role within these computations.

The role of consciousness is so obscure that many authors doubt it, considering phenomenal consciousness as an *epiphenomenon*. Epiphenomena are known in evolutionary biology. Features which were not selected for, but result from the selection of other characteristics, are evolutionary epiphenomena. The most often mentioned example is the human chin, which appeared as a consequence of face and jaw reduction. The chin is not an organ shaped by evolution in the first place. Similarly, if consciousness is considered as a mere property accompanying some neural mechanisms, it is nothing more than a fortuitous by-product of brain evolution. Any evolutionary epiphenomenon has two basic properties : it is fortuitous and neutral. It could have been different or non-existent, and it has no effect on the survival of individuals. Is consciousness such a fortuitous, neutral feature ? In our view, the fact that phenomenal consciousness is systematically associated to sensory input analysis indicates that it is not incidental.

Phenomenal consciousness may be considered as an epiphenomenon in another sense. It is sometimes said to be an emergent feature of complex functional organisations. This concept of consciousness emerging from complexity is however not operational. It does not explain why every brain region does not equally contribute to consciousness [Edelman 1989]. It does not explain either why brain damage may alter phenomenal experience selectively.

A very serious claim is that phenomenal consciousness is systematically associated to a given physical neural state [Edelman 1989 ; Damasio 1989 ; Crick & Koch 1990] :

> Our basic hypothesis at the neural level is that it is useful to think of consciousness as being correlated with a special type of activity of perhaps a subset of neurones in the cortical system. Consciousness can undoubtedly take different forms, depending on which parts of the cortex are involved, but we hypothesize that there is one basic mechanism (or a few) underlying them all [Crick & Koch 1990, p. 266].

These authors consider consciousness as an authentic biological feature, but nothing prevents us from putting forward such a hypothesis to depict consciousness again as an epiphenomenon : what was retained by selection would not be consciousness itself, but the underlying neural mechanism. In this kind of description, consciousness plays no causal role by itself in cognitive activity. It is not supposed to be a mere fortuitous side-effect, it is the mental correlate, experienced from a personal point of view [Nagel 1974], of a special kind of brain activity.

Our claim about the modular properties of qualitative experience will allow us to argue against epiphenomenon hypotheses and to put forward a possible role for phenomenal consciousness in evolution.

*2.2 Modularity of qualitative experience*

The existence of conscious experience, which has recently become the object of many scientific and philosophical investigations, seems to deserve closer examination. The quality of sensory states at the phenomenal level – how things look, sound, how we feel them – appears to be modality-dependant. Mental disorders occurring after brain injuries sometimes reveal that some particular aspect of consciousness may be selectively impaired. For instance blind-sight patients declare to be blind in a certain area of their visual field. These patients suffer from brain damage, and their blind area corresponds precisely to the location of lesions in the primary visual cortex. However, it has been shown that they are still able to perform visual processing like localising simple visual stimuli, elementary patterns or movements [Weiskrantz 1980, 1987]. These patients are totally unaware of their residual visual capacity. They just claim they are "guessing" during visual tests. Their phenomenal experience is selectively impaired in the visual modality.

Different types of neuropsychological syndromes (like amnesia, hemineglect, agnosia) that alter or suppress aspects of qualitative experience suggest the existence of dissociations within the sensory domain of information processing. As far as we can conclude from such neural deficits, each property of a given experience seems to be produced by a fixed and specialised neural architecture. These highly selective syndromes suggest that phenomenal consciousness is not globally distributed, but modular and that its modular properties mirror the organisation of sensory input modules.

Qualitative aspects of experience originate at the output of sensory modules[2]. They are and remain separate (we never confuse the redness of an apple with its taste). Memory and perception are never experienced as a mixture of indistinct sensations. Qualitative experience is also mandatory : you can't avoid experiencing redness when you look at a red screen, seeing a visual array as a three-dimensional objet, or hearing an utterance of a sentence (in a language one knows) as a sentence.

These modular properties suggest that an adaptive role for phenomenal consciousness is to improve the ability to discriminate perceptual and mental states.

## 3. Qualitative labelling of experience

*3.1 Cognitive labelling through phenomenal properties*

One of the most basic and important tasks a living creature has to perform in order to eat, move, mate and avoid predation is to extract relevant information from its sensory inputs and from its memory. This is what information processing is all about. The task is indeed not a trivial one. Biologically relevant information is indirectly defined by genes, possibly through learning, though genes can only give a rough indication. This is sometimes sufficient. For a frog, any small flying object is *a priori* edible. For learning to take place, however, situations must be distinguished. A frog is unable to learn anything about flies, since all flies look the same. With a specialised device for labelling experienced situations and a simple feed-back like edible / non-edible, a creature can learn a lot about things relevant to food.

There are thus two basic labelling functions : (1) evaluation, in order to mark situations as positive or negative according to various scales (edible, dangerous, attractive, etc.) ; (2) perceptual labelling, which aims at individualising contexts for complex representational processing. Our claim is that phenomenal consciousness performs both functions, and that this is its main biological purpose, the reason for which it has been selected during phylogenetic history.

The requirements are different for each label type. Perceptual labels are necessarily the result of a highly combinatory device, so that many distinct labels can be generated, while a " value " label must have a wide dynamic range, so that such labels can be accurately compared (the two requirements are not mutually exclusive). Our ability to simultaneously experience colour, shape, temperature, weight, sound features, distance and so on meets the first requirement. Any combination of all such parameters is likely to be unique. On the other hand, our experience of physical pain or pleasure, of sadness or joy, of pride, of nostalgia, etc. ranges from slight feeling to extreme intensity, and is thus suitable for comparative assessments of situation significance.

The perceptual labelling role we give to phenomenal consciousness can be inferred from psychological studies on memory source monitoring. Johnson et al. [1988 ; 1993] claim that the phenomenal qualitative properties of mental experiences are the very source of a more general process of discrimination, judgement and attribution of mental events. They suggest that phenomenal properties of experience play a critical role in discriminating knowing from remembering, and thus, create a source for one's sense of personal past. When memory information without qualitative characteristics is recalled, it is experienced as mere knowledge or belief. Hence, phenomenal properties relating to perceptual and contextual information appear as important cues for discriminating representations, which is essential for assessing the reliability of information. Confusion about the nature and the source of different mental representations is likely to be the cause of misattribution in the confabulation syndromes where amnesiac patients actually mix up the imagined, perceptual and memory representations.

From an engineering perspective, a modular labelling system appears to be rather odd. Labelling information is indeed a simple task in computer science and data transmission. All you have to do is to add unambiguous labels, *e.g.* as headers, to messages. When you receive data on your network navigator or when you open a file with a word processor, objects received or read identify themselves as text, picture, table or whatever because they contain heading information giving their type. Labelling on a computer is thus easily performed by adding information to information. However, there is an obvious difference between computers and neural networks, so obvious that it remains implicit and is sometimes overlooked. This difference presumably prevented evolution from using headers as labels. To put it crudely, information is never merely transmitted in a neural network. Let us briefly clarify this point.

By definition, in a digital communication context, a message becomes information only when appropriate operations have picked out its features. This definition fits in with what we know of sensory analysis in the mammal brain. Take the example of an image. When received on a retina, it is a mere matrix of pixels, bearing no information in terms of boundaries or outlines. This latter information appears after it has been detected by edge detectors. What is transmitted to a further processing level, *e.g.* an object recognition level, is expressed in terms of lines or edges, no longer in terms of pixels. In such a processing sequence, information is never transmitted as such, because at each stage, the symbol set changes. Things are different on a computer : you may mark out a piece of text to indicate the make-up and still have a text, mostly composed of the same characters with a few additional marks. In neural networks, this is impossible. Any processing changes the nature of information[3]. An edge detector is fed with pixels, but its output is of a different kind : it indicates the presence or absence of an edge with a given orientation at a given location. The engineering solution which consists in adding headers to a message in order to identify it unambiguously, as for instance in electronic mail, does not work with neural circuitry[4], since such headers would be lost at each processing stage.

One possibility is to consider that perceptive details have to be forgotten at higher levels of a hierarchical cognitive architecture. Detailed features play a role at the first stages of recognition, but are of no use afterwards when abstract features are processed. C. von der Malsburg [1986] shows in detail why such an organisation is not convincing at all, because of its lack of flexibility and parsimony. A purely hierarchical system necessitates the existence of dedicated units to represent high-level patterns. But whereas the number of combinations that have to be distinguished is virtually infinite, the number of such dedicated units in the brain are certainly limited. In other words, a purely hierarchical organisation lacks combinatorial power. Also, such dedicated units being separate, they cannot be the basis for generalisation :

> When I consider a particular scene, I absorb knowledge about the objects involved, by modifying the interactions within and between the corresponding mental symbols. I want to be able to have this knowledge at my disposal in other situations if they involve partly the same objects or aspects. This, however, is possible only through physical overlap between mental symbols. Avoiding this overlap destroys the basis for generalization. [von der Malsburg 1986]

Von der Malsburg suggests that "mental symbols" are not limited to a given hierarchical processing level. What he calls the "natural representation" of an object simultaneously involves all its constituent elements. As a consequence, no information is lost in the integration process.

> The symbols of communication [*e.g.* written words] are mere parsimonious tokens for the images they are to evoke in the reader's mind. In contrast, the symbols of mind have to fully represent all aspects of our imaginations. [von der Malsburg 1986]

If we accept this kind of description, phenomenal qualities, which appear at the output of modular sensory systems, are available for higher-level processes. In this context, the labelling role played by phenomenal consciousness becomes manifest. Processes like the justification and the revision of beliefs, especially perceptual beliefs, are sensitive to qualitative aspects that are only present at non-conceptual levels. For instance, in order for such operations to be accomplished, one should preserve the *origin* (*e.g.* external *vs.* internal) of the representations which are poised for use in reasoning and in the rational control of speech and action. The perceptual origin of the representation seems to be assessed from the abundance of phenomenal details. Kelly and Jacoby [1993] argue that the feeling of familiarity arises from attributions based on internal cues, such as the ease or relative fluency of perceptual operation, the quality of memories and the vividness of visual images. The experience of remembering is not the result of some intrinsic qualities of "memory trace", but rather reflects the operation of a decision process that assigns ongoing mental events to particular sources. People normally use the presence of perceptual details in a mental state as a cue to infer that they are recalling, rather than imagining, and perceiving rather than remembering. Phenomenal and qualitative properties accompanying some kinds of mental states, *e.g.* perceptual or proprioceptual states and some episodic memory states, are important cues that enable us to ascribe them to ourselves[5,6].

Phenomenal qualities seem thus to strongly interfere with higher-order cognitive processes. As a consequence, we are always in hybrid mental states, partly conceptual and partly made of contextual qualitative information. The labelling of conceptual representations by qualitative properties is only possible if the latter may enter as constituents in cognitive representations, as suggested by von der Malsburg. Recent advances in brain modelling makes this requirement plausible, as described below.

*3.2 Neural labelling implementation*

With phenomenal consciousness, natural selection seems to have discovered a way of labelling inputs which is compatible with neural implementation. But how is it implemented ? Edelman [1989] suggests that conscious perception relies on active categorisation. He explains that a set of several neural maps is responsible for the integrated conscious perception of scenes. This set of maps has been selected among other possible combinations of groups of neurones during ontogenesis. Perception itself results from the selection of a neural circuit among all possible combinations of connections between maps, through a process called *reentry*, which is a recurrent exchange of signals between maps. This is supposed to explain how sensory input analysis can be distributed over several locations in the brain and still produce a unified perception that is rich enough to be categorised. Thanks to reentry, perception is compared with memory traces through an active process that modifies both perception and memory. Primary consciousness results from these categorisation processes. Edelman, using the same principle, explains how such a unified, conscious, perception of a scene is connected to what he calls "values". Reentry is supposed to occur between cortical maps and specific locations in the limbic system that implement values. The latter connection accounts for the evaluation of the perceived situation. Areas responsible for evaluation (esp. limbic system, hypothalamus, brain stem) are phylogenetically older than those performing categorisation (thalamus and cortex). Both systems are necessary for consciousness.

This account by Edelman is attractive, but it is far from being fully developed. For instance, Edelman's theory does not help understand why some complex cognitive processes are performed unconsciously. Also, Edelman's description is a purely neuronal account. There is no indication of any specific role that qualitative properties of experience could play, even if the author claims that consciousness is cognitively efficient and increases evolutionary adaptation of individuals. We shall now consider another neural account of phenomenal consciousness that may allow us to avoid these drawbacks.

Our hypothesis is that phenomenal consciousness has an adaptive function which is to allow discrimination and labelling of perceptual and mental states. The issue of knowing how labelling is achieved is connected to a problem concerning perception itself, known as the *binding problem*. As Damasio puts it :

> It is not enough for the brain to analyze the world into its components parts : the brain must bind together those parts that make whole entities and events, both for recognition and recall. Consciousness must necessarily be based on the mechanisms that perform the binding. [Damasio 1989]

In the brain, contrary to what happens in computers, different kinds of processing occur in different locations. For instance, colour analysis, shape recognition, movement and several other characteristics of visual scenes are detected in separate parts of the visual cortex. However, our brain constructs a single and global view of the scene. This integration requires a binding mechanism, so that we are able to simultaneously assign red colour, direction and form to a single object of the visual scene, that object moving toward us over there that we identified as a car. Objects exists as complex representations in our mind because we are able to link several phenomenal characteristics we could extract from our sensory processing and correlate them together as single objects. As we said, qualitative experience is not a general property of our mental states and mental processes. We claimed that different aspects of experience depend on different sensory modalities. However, qualitative properties experienced in a given situation are bound together across modalities and are unified into a single representation.

Synchronous neural activity, since von der Malsburg [1986] and others, is often invoked to account for binding. It has been experimentally observed that neurones located in different cortical areas may function synchronously [Singer 1993]. Evidence from neurophysiology and from connectionist studies [von der Malsburg & Schneider 1986] suggested that frequency locking between neurone groups could account for the integration of different features of a given perceived situation.

Binding through synchronous neural activity is temporary. This explains why its combinatory power is virtually infinite. As Singer [1993] puts it, "the essential advantage of assembly coding is that individual cells can participate at different times in the representation of different objects". Hence every combination of extracted characteristics can be integrated into a single representation and possibly memorised as such. This combinatory power is what is needed for a perceptual labelling device. Our suggestion is thus that (1) dynamic feature binding allows labelling of situations ; creatures with this ability can cope with much more complex environments ; (2) phenomenal consciousness was selected as a way to perform labelling through binding.

At this point, we have an idea about the kind of adaptive role played by phenomenal consciousness. We also have plausible models of the way the labelling function may be implemented. We still need accounts for the role phenomenal consciousness played in its own evolutionary emergence. Was it directly selected, or is it an evolutionary epiphenomenon ?

## 4. An evolutionary role for phenomenal consciousness

*4.1 Phenomenal variety and signal discrimination*

The claim that qualitative experience directly contributed to the ability of individuals to adapt to their environment during phylogenesis is equivalent to saying that qualitative experience is part of the phenotype. In evolutionary systems, we call phenotype the set of characteristics which are directly evaluated in the selection process [Dessalles 1992,1996]. Let us consider an analogy. Ethologists consider bird songs as adaptive : a mute song bird would not perform well, being unable to signal its territory properly. The ancestors of song birds were selected for their ability to sing. Should we consider that singing itself was selected, or rather that the syrinx (bird pharynx) was selected in order to allow territory signalling ? Perhaps we should look at the neural processes that are involved in singing and say they were also selected for territorial signalling purposes. What did selection retain after all, if not the genetic changes that make the difference between song birds and their non-singing ancestors ? From genes to neural processes, syrinx and song, there is a long chain of embryological events. Each of them is necessary for singing to occur. However, when ethologists study song birds, they are more prone to consider that the song itself was shaped by evolution to perform territory signalling, rather than syrinx or neural states. There are two reasons for this : first, actual songs seem to be optimal according to the way "fitness" (here efficient territory signalling) is assessed[7] ; second, the fitness of the song can be assessed directly, whereas the fitness of syrinx is indirect and we must refer to the singing ability[8].

For the same reasons, we claim that from an evolutionary perspective we should include phenomenal consciousness into the phenotype of conscious beings rather than the neural states that underlie qualitative experiences. We indicated how phenomenal consciousness, through its labelling ability, could be assigned a fitness value. Now we want to show that qualitative properties of experience are, in a sense, optimal for the labelling ability. We should however be aware of two difficulties. Bird song can in no way be considered as an evolutionary epiphenomenon as phenomenal consciousness can. Also, even if song is a more abstract entity than physiological organs, it can be objectively measured, whereas qualitative experiences are not accessible : they are private to a single, subjective perspective [Nagel 1974].

We assume that phenomenal consciousness is a biological characteristic of living species, so we should be able to account for it in terms of natural selection. Any observed complex characteristic of living beings which is not a side-effect must have (or have had) an adaptive value[9]. We suggested that phenomenal consciousness is associated with an adaptive function, which is to label experience at the output of perceptual systems, in such a way that representations do not necessarily become purely abstract when they reach central systems. However, we have no direct evidence showing that phenomenal consciousness was itself selected to perform this labelling function. We still have to discard the possibility that it is an evolutionary epiphenomenon : neural processes could have been selected directly to perform the labelling function, and they would happen to have phenomenal correlates. The question is thus to know whether qualitative experiences are phenotypic or not. Can we assess the optimality of neural processes performing labelling without making reference to qualitative experiences ?

We want to suggest that there is a "mapping" between the physical input space and the qualitative space, and that such a mapping is not predicted by the epiphenomenon hypothesis. Consider an example from phonology. The three vowels [a], [i] and [u], present in words like *apple*, *see*, and *fool*[10], are basic phonemes present in virtually all natural languages [Maddieson 1984]. Being able to distinguish them is thus essential for any human being. [a], [i] and [u] look indeed very different to a human ear. This qualitative contrasted appearance is consistent with the fact that the discrimination performance is maximum for these vocalic phonemes [Lindblom 1986]. It can be shown though spectral analysis that these three phonemes are objectively "distant" : by measuring basic spectral characteristics called "formants", acousticians show that [a], [i] and [u] are located in opposite corners of the accessible space. These studies by acoustic engineers are generally considered as relevant because they establish an objective link between our intuition (the three vowels look different) and the requirements of robust communication (symbols used for communication should be maximally different to be easily distinguished). From another perspective, however, such an apparently plausible result should be regarded as quite unlikely. Why should our qualitative feeling about the dissimilarity of these phonemes be correlated with communication requirements ? If qualitative experience is nothing but an evolutionary epiphenomenon, we would expect no such dissimilarity between qualitative states corresponding to the perception of [a], [i] and [u].

This example reminds us that for some discrimination tasks[11], it seems that we are fully aware of all the differences we are able to detect. In other words, in such cases, our discriminatory power is entirely due to the grain of qualitative conscious aspects of our experience. Our performance relies on the fact that all the qualities we are able to experience in a given modality are different and separate. We can take other examples involving colour or flavour discrimination. We are aware of all colour shades that we can discriminate. This good performance, compared to other mammal species, is due to the fact that normal human beings[12] experience different wavelengths in a contrasted way. For instance, colours usually distinguished in English have quite contrasting qualitative appearances. We can even assess subjective distances by saying that blue is closer to violet than to yellow. Similarly, pineapple taste is not so far from lemon, but not at all like tomato. All the stimuli which are biologically relevant and that we effortlessly discriminate induce clearly distinct qualitative experiences. This is hard to explain if phenomenal consciousness was not involved in the evolutionary process. Why aren't there colours (or tastes or sounds) that we would experience as identical but that we would still be able to discriminate ? If phenomenal experience was a mere by-product of neural evolution, we could suppose that only neural processes are needed for detecting physical information without calling for the corresponding qualitative states.

Phenomenal variety, the fact that qualitative experiences in a given modality are differentiated, may be given a technical explanation. It is well-known, from an engineering perspective, that signal discrimination is easier if signals are spread over a wide energy range and compared to maximally distinct patterns[13]. [a], [i] and [u] are acoustically the most distinctive vocalic sounds our vocal tract is able to emit. The fact that we experience phonemes like [a], [i] and [u] as clearly distinct suggest that phenomenal properties are involved in the discrimination process and that they carry information.

The only possibility which is consistent with phenomenal variety is that qualitative experience is *not* an evolutionary epiphenomenon : it plays a direct role in discrimination and as a consequence was selected for its own sake. In other words, we perform discriminations on the basis of phenomenal qualities. First conscious species were selected according to this ability which requires a rich repertoire of phenomenal qualities in each modality. The fact that qualitative experience has a modular structure that systematically mirrors the organisation of perceptual systems, and the fact that it meets constraints of signal discrimination efficiency by keeping relevant qualitative properties scrupulously apart, suggest that phenomenal consciousness was itself involved in the evolution process.

*4.2 Selection pressure on qualitative experience*

Our claim is that phenomenal consciousness is optimally designed to perform its function, which in our view is to label perceptual and mental states. It is associated with the output of each modal sensory processing where it makes relevant signals the most discernible. This is exactly what we expect from a perceptual labelling device designed by natural selection. If we accept this hypothesis and think that phenomenal consciousness has been directly produced by evolution to fulfil an adaptive function, then we may consider (1) that phenomenal consciousness *is phenotypic* and (2) that neural states underlying phenomenal states only exist because the latter have an adaptive function. In this sense, *phenomenal consciousness is part of the phenotype*, exactly as bird song in our example. Underlying neural devices are not themselves phenotypic, since they are just a link in the long chain going from genes to phenomenal consciousness. If we follow the analogy with bird song, optimality of qualitative experience can be directly understood, whereas the optimality of underlying neural states would only appear through a reference to phenomenal properties.

From this perspective, phenomenal consciousness is what led the evolution of cognitive systems towards increasing discriminatory capacities. If phenomenal qualities were epiphenomenal, our perceptions would not give rise to such a variety of phenomenal states. The richness and the vividness of our phenomenal repertoire suggests that it is the direct product of natural selection. Under this hypothesis, qualitative experience has to be seen as a driving element in the evolutionary process which produced both our rich perception of the environment and our ability to discriminate mental states. It is thus indirectly responsible for our ability to learn efficiently.

## 5. Conclusion

We presented phenomenal consciousness as modular. Qualitative properties of experience are associated with sensory modalities, they are and remain distinct even if they can be integrated into multimodal and conceptual representations of objects and events. According to the hypothesis presented here, an adaptive function of phenomenal consciousness is to be found in relation to this integration involving qualitative information. Qualitative properties play the role of labels. Through the combinatorial power of a binding mechanism based on synchronous firing of neurones, representations may be multimodal and yet preserve contextual and modally distinguished perceptual aspects. Conscious organisms are thus able to discriminate among their perceptual representations. They are neither highly specialised robots nor purely abstract general problem solvers. Phenomenal consciousness allow them to better cope with the wide range of situations found in a complex ecological environment.

Higher-order cognitive processes have to be sensitive to qualitative properties of experience in order to determine the source of mental representations. According to our hypothesis, this is made possible by the fact that qualitative properties play the role of labels that carry information about the origin of representations.

The structural features of phenomenal consciousness, its modularity and the variety of qualitative properties within each modality, are in accordance with what we expect from a labelling device. On the other hand, alternative accounts in terms of neural states that consider qualitative properties as epiphenomenal can hardly explain the richness and the vividness of the qualitative repertoire. Phenomenal consciousness should be considered as a proper phenotypic character. *Phenomenal consciousness is what natural selection could act upon*. Any increase in qualitative variety was likely to induce a more probable survival of individuals. This might explain why phenomenal properties of experience, which seem to be optimally designed for the labelling of representations, were selected and designed by evolution.

## 6. Notes

[1] The feeling of being a single entity, the fact that some recalled events look familiar, the feeling of "ownership" about our mental states, the first-person point of view, the ability to observe aspects of our cognitive functioning are other important features of what is called consciousness. Nevertheless, all of them are different aspects of consciousness, each one might be related to different cognitive functions and may eventually call for different accounts [Zalla 1996].

[2] In the modular theory of consciousness put forward by R. Jackendoff [1987], only the intermediate level, where sensory information has been processed in a modality specific way but has not yet reached central representations, supports awareness.

[3] The reader may object that topological information is transmitted as such, from map to map, in neural visual processing. But what is conveyed here is signal, not information. Neighbouring relations are present in the matrix for an external observer, but they do not exist as such for the brain until they are detected. And they are lost afterwards. An edge detector may use topology among pixels. At the output of this detector, topological relationships between pixels do not exist anymore, simply because at this stage pixels are no longer represented. Topology among edges is preserved in the signal, only because it has not yet been detected.

[4] We speak here of biological plausible circuitry as we imagine it, since it is technically possible to perform anything with neurones, even compute square roots.

[5] Suengas and Johnson's experiments [1988, p.388] also demonstrated that both emotion when recalling imaginary events and lack of clarity when recalling real events reduce qualitative differences between these two types of memories and thus tend to generate some kind of confabulation.

[6] A syndrome associated with deep lesions in the right posterior, non-linguistic hemisphere is characterised by the patient's denial of "ownership" of his paralysed, left arm. Conversely, normal subjects experience the loss of a limb very much as a loss of "a part of themselves". We can suppose that the lack of proprioceptual qualitative states is the cause of one's misattribution of parts of the body.

[7] For instance, characteristics of bird songs produced by different neighbouring species are very different. The male bird can thus be correctly identified by females of its species.

[8] By contrast, a physiologist would not be interested in territory signalling. She would consider syrinx as phenotypic and the ability to produce a distinctive song as a way to assess syrinx fitness.

[9] Strictly speaking, the adaptive value should be assessed at the gene level [Dawkins 1978]. Neutralists [Kimura 1983] have claimed that random shifts are an important aspect to explain evolution ; however the probability that complex functional characteristics emerge from random shift is virtually zero.

[10] In French, these phonemes are present in words like *plat*, *vie*, *roue*. In English, *apple* starts with [æ]. Better examples for [a] would be words like *lie* and *now* in which the first part of the diphthong is considered.

[11] According to the modular description that we adopted, this happens at a certain level of input analysis, at the output of sensory modules.

[12] Colour blind subjects being of course excluded.

[13] In digital communications, possible waveforms should be chosen so that the energy of their difference is maximal.